\documentclass[prc,aps,nofootinbib,%
tightenlines,superscriptaddress,floatfix,twocolumn,%
preprintnumbers]{revtex4}
\usepackage{graphicx}
\usepackage{amsmath}
\usepackage{amsfonts,amsbsy}
\usepackage{amssymb}

\def\empile#1\over#2{\mathrel{\mathop{\kern 0pt#1}\limits_{#2}}}

\def\gsim{ \,\, \vcenter{\hbox{$\buildrel{\displaystyle >}\over\sim$}}
 \,\,}

\def\qp{ {\bf q}_\perp } 
\def\pp{ {\bf p}_\perp } 

\def\kp{ {\bf k}_\perp }

\newcommand{\ud}{\mathrm{d}}

\catcode`\@=11

\newcount\@tempcntc
\def\@citex[#1]#2{\if@filesw\immediate\write\@auxout{\string\citation{#2}}\fi
  \@tempcnta\z@\@tempcntb\m@ne\def\@citea{}\@cite{%
        \@for\@citeb:=#2\do%
    {\@ifundefined{b@\@citeb}%
        {\@citeo\@tempcntb\m@ne\@citea%
                \def\@citea{,\penalty\@m\ }{\bf ?}\@warning%
                {Citation `\@citeb' on page \thepage \space undefined}}%
        {\setbox\z@\hbox{\global\@tempcntc0\csname b@\@citeb\endcsname\relax}
     \ifnum\@tempcntc=\z@ \@citeo\@tempcntb\m@ne%
       \@citea\def\@citea{,\penalty\@m}%
       \hbox{\csname b@\@citeb\endcsname}%
     \else%
      \advance\@tempcntb\@ne%
      \ifnum\@tempcntb=\@tempcntc%
      \else\advance\@tempcntb\m@ne\@citeo%
      \@tempcnta\@tempcntc\@tempcntb\@tempcntc\fi\fi}}\@citeo}{#1}}%

\def\@citeo{\ifnum\@tempcnta>\@tempcntb\else\@citea
  \def\@citea{,\penalty\@m}%
  \ifnum\@tempcnta=\@tempcntb\the\@tempcnta\else
   {\advance\@tempcnta\@ne\ifnum\@tempcnta=\@tempcntb \else
\def\@citea{--}\fi
    \advance\@tempcnta\m@ne\the\@tempcnta\@citea\the\@tempcntb}\fi\fi}

\catcode`\@=12

\begin{document}

\title{The ridge in proton-proton collisions at the LHC}
\author{Adrian Dumitru}
\affiliation{RIKEN BNL Research Center, Brookhaven National Laboratory,
Upton, NY-11973, USA
}
\affiliation{Department of Natural Sciences, Baruch College,
CUNY, 17 Lexington Avenue, New York, NY 10010, USA
}
\author{Kevin Dusling}
\affiliation{Physics Department,   Brookhaven National Laboratory,
  Upton, NY-11973, USA
}
\author{Fran\c cois Gelis}
\affiliation{Institut de Physique Th\'eorique (URA 2306 du CNRS),  CEA/DSM/Saclay, 
  91191, Gif-sur-Yvette Cedex, France
}
\author{\mbox{Jamal Jalilian-Marian}}
\affiliation{Department of Natural Sciences, Baruch College,
CUNY, 17 Lexington Avenue, New York, NY 10010, USA
}
\author{Tuomas Lappi}
\affiliation{Department of Physics, P.O.~Box 35,
  40014 University of Jyv\"askyl\"a, Finland
}
\affiliation{Helsinki Institute of Physics, P.O.~Box 64,
00014 University of Helsinki, Finland
}
\author{Raju Venugopalan}
\affiliation{Physics Department,   Brookhaven National Laboratory,
  Upton, NY-11973, USA
}

\begin{abstract}
  We show that the key features of the CMS result on the ridge
  correlation seen for high multiplicity events in $\sqrt{s}=7$~TeV
  proton-proton collisions at the LHC can be understood in the Color
  Glass Condensate framework of high energy QCD. The same formalism
  underlies the explanation of the ridge events seen in A+A collisions
  at RHIC, albeit it is likely that flow effects may enhance the
  magnitude of the signal in the latter.
\end{abstract}

\preprint{INT-PUB-10-051}
\preprint{BCCUNY-HEP/10-03}
\preprint{BNL-94103-2010-JA}
\preprint{RBRC-858}

\maketitle

\section{Introduction}
\begin{figure}[bht]
\begin{center}
\includegraphics[width=0.48\textwidth]{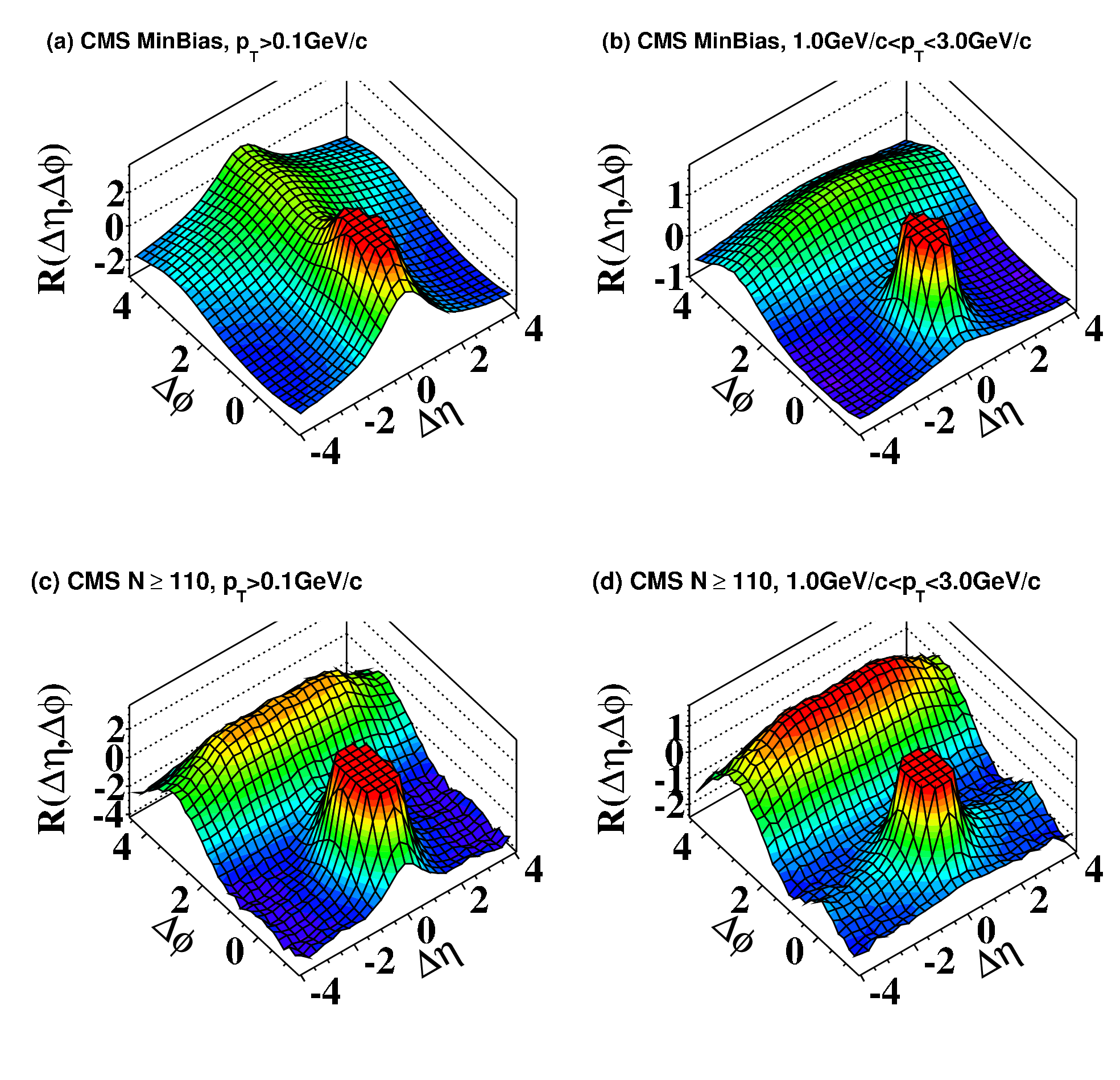}
\includegraphics[width=0.48\textwidth]{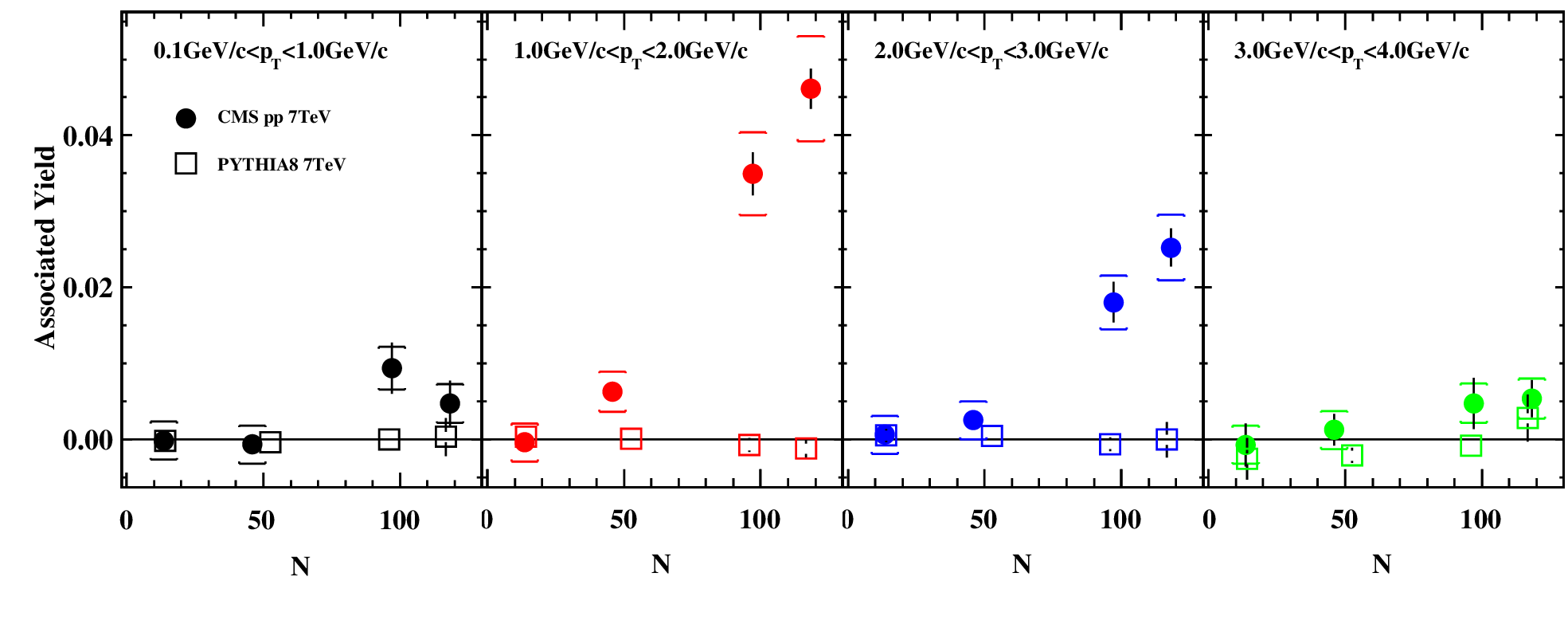}
\end{center}
\caption{\label{fig:one}Top: 3-D display of the two particle
  correlation $R$ as a function of $\Delta\eta$ and $\Delta\phi$ for
  minimum bias and high multiplicity events in two different $p_\perp$
  windows. Bottom: The associated yield in different $p_\perp$ windows
  as a function of the number of charged particle tracks. From
  ref.~\protect\cite{CMS1}.}
\end{figure}

A very recent preprint~\cite{CMS1} from the CMS collaboration at the
LHC reports the observation of a ridge-like structure of correlated
charged particle pairs with momenta in the range $p_\perp,q_\perp \sim
1$--$3$~GeV in high multiplicity (with $N \geq 110$) events in
proton-proton collisions at $\sqrt{s}=7$~TeV. This ridge is a feature
on the ``near side'' of the two particle correlation, around $\Delta
\phi \approx 0$ azimuthal separation between the two particles. It
extends at least up to $\Delta \eta \approx 4.8$, the limit of
acceptance of the detector components used in the analysis. The ridge
is seen only in the moderate $p_\perp,q_\perp$ range and
systematically vanishes for $p_\perp,q_\perp \lesssim 1$~GeV and
$p_\perp,q_\perp \gtrsim 3$~GeV.  A plot summarizing this structure in
the CMS data is shown in fig.~\ref{fig:one}.

\begin{figure}
\begin{center}
\includegraphics[width=0.45\textwidth]{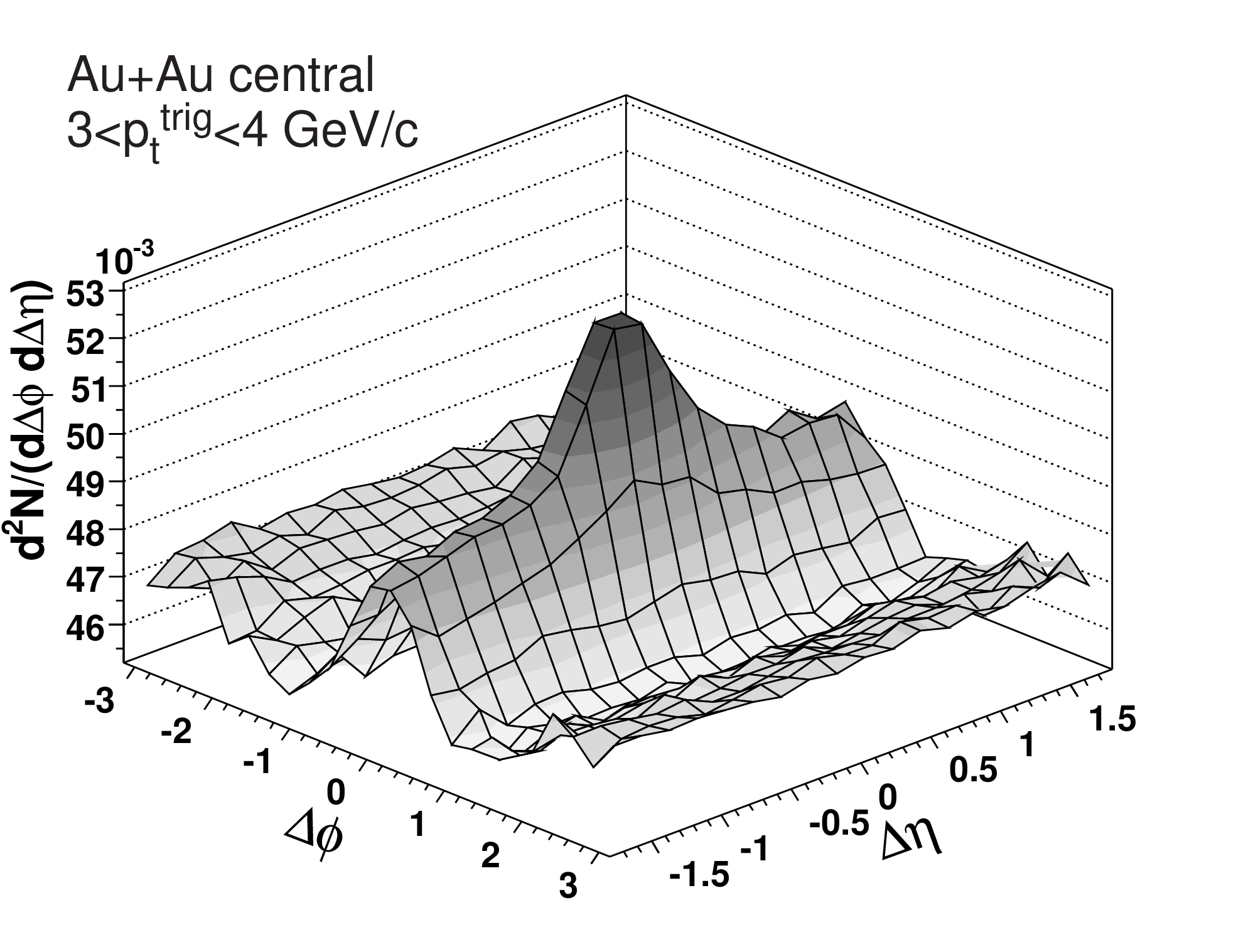}
\includegraphics[width=0.45\textwidth]{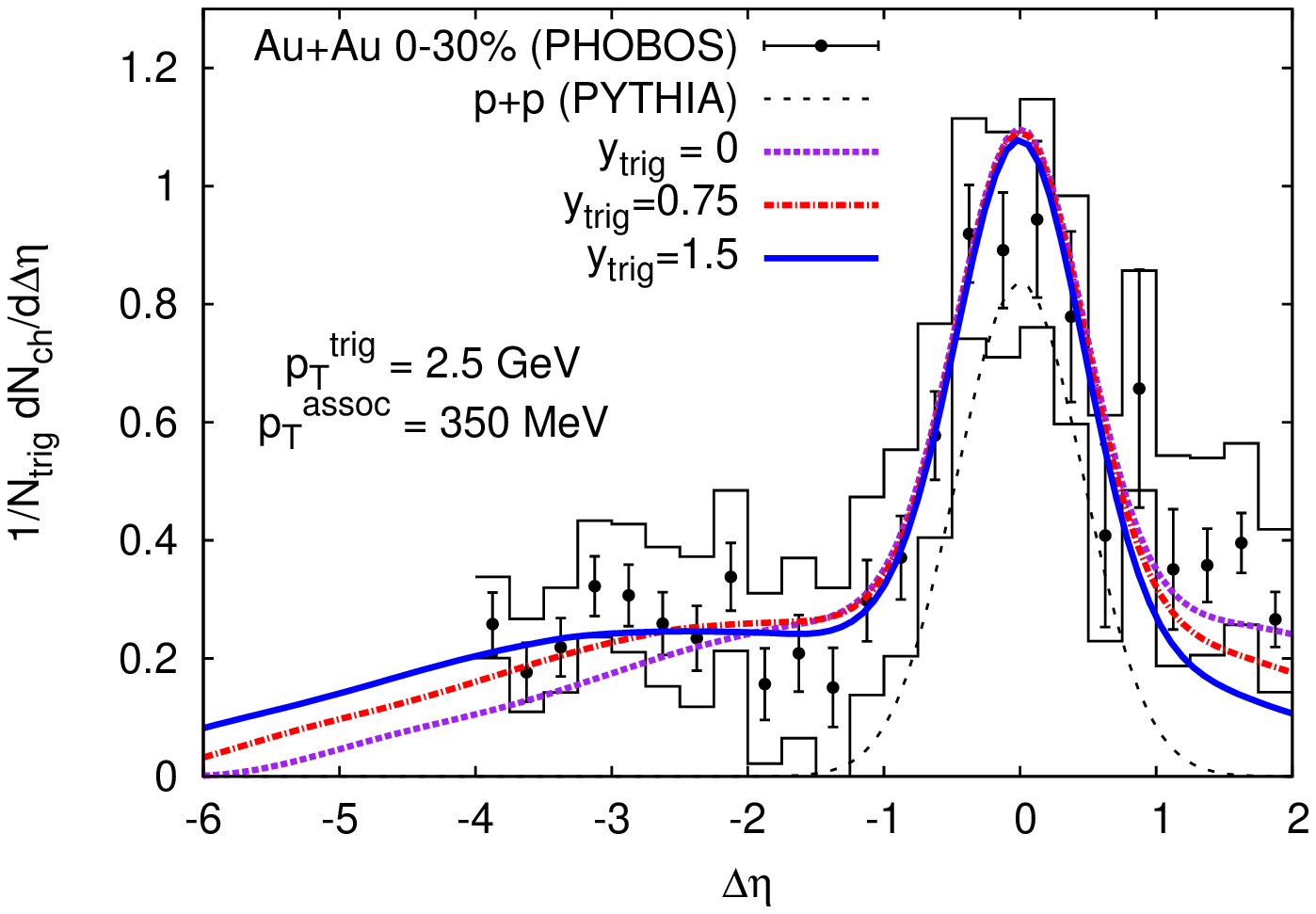}
\end{center}
\caption{Top: 3-D display of the two particle correlation
  (analogous to $R$ in fig.~\ref{fig:one}) for
  $p_\perp$-triggered events from the STAR
  collaboration~\cite{Abelev:2009qa} (note that the 
  away-side peak  around $\Delta \phi = \pi$  
  seen in fig.~\ref{fig:one} has been removed). 
  Bottom: Two-particle correlation data
  from the PHOBOS collaboration~\cite{AlverA1} that reveals a long
  range component. The curves shown are obtained by adding our result
  (in eq.~(\ref{eq:double-inclusive})) to the short range correlation
  from PYTHIA.  }
\label{fig:two}
\end{figure}

A similar ridge-like structure was previously seen in nucleus--nucleus
collisions at
RHIC~\cite{Adamsa4,Adamsa5,Adarea1,AlverA1,Abelev:2009qa,Alver:2010rt}. The ridge
was seen in high multiplicity (central) events in Cu+Cu collisions at
$\sqrt{s}=62.4$~GeV and in Au+Au collisions at $\sqrt{s}=200$~GeV. The
STAR detector observed this correlation for both
$p_\perp$-triggered~\cite{Adamsa4,Abelev:2009qa} and untriggered~\cite{Adamsa5} pair
correlations in the whole STAR TPC acceptance of $\Delta \eta \leq 2$.
The PHOBOS experiment~\cite{AlverA1} observed the $p_\perp$-triggered
correlation at much larger rapidity separations, initially up to
$\Delta \eta\sim 4$, extended more recently~\cite{Alver:2010rt} to
$\Delta\eta \sim 5$.  The long range correlation structure disappears
for lower multiplicity peripheral events in nucleus-nucleus collisions
and are also absent in deuteron-gold and proton-proton ``control''
experiments at the same energies. The STAR and PHOBOS correlation
results are shown in fig.~\ref{fig:two}.

\begin{figure}[tbh]
\begin{center}
\includegraphics[width=0.48\textwidth]{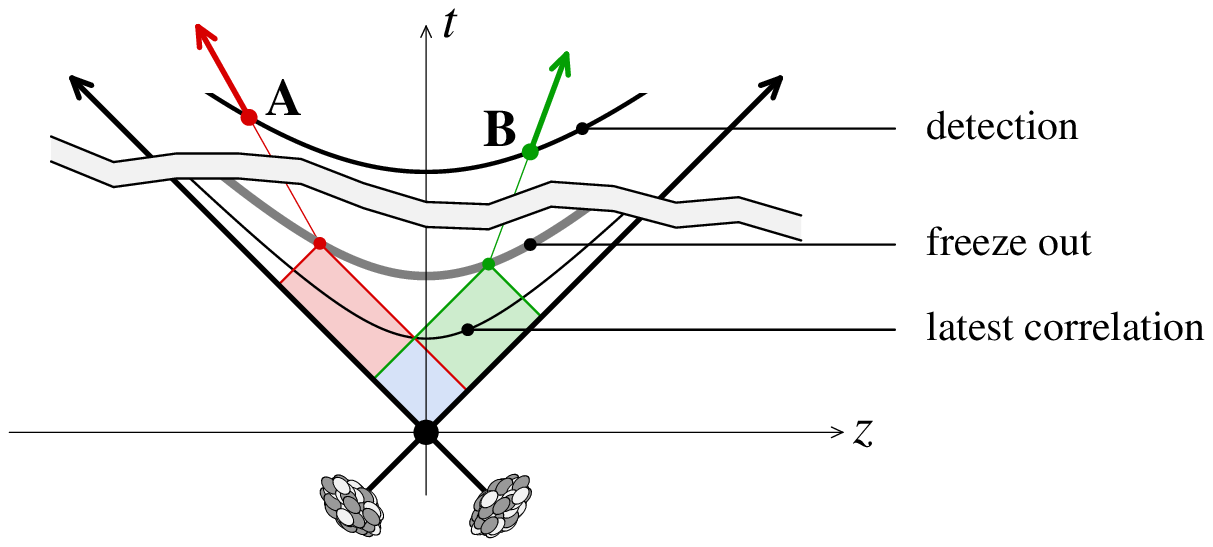}
\end{center}
\caption{This figure illustrates the argument from causality that long
  range correlations of particles (denoted $A$ and $B$ here) must
  occur at very early proper times. The doubly shaded region,
  corresponding to the intersection of the light cones of the two
  particles, is the space-time location where correlations can be formed.}
\label{fig:three}
\end{figure}

Why is the ridge interesting? One can show from a simple argument
based on causality~\cite{DumitGMV1} that if long range rapidity
correlations between particles exist, the correlation must be formed
at proper times earlier than\footnote{This argument assumes that a
  produced particle's momentum space  rapidity 
  is tightly correlated with its space-time rapidity.
}
\begin{equation}
\tau_\textrm{init.} = \tau_{\rm f.o.}\, \exp\left(-\frac{1}{2} \Delta y\right) \,,
\end{equation}
where $\tau_{\rm f.o}$ is the freeze-out time of the particles. This
causality argument is illustrated in fig.~\ref{fig:three}.  If we
assume this freeze-out time to be of the order of a few fermis, a
rapidity correlation of $\Delta \eta \sim 5$ units suggests that these
correlations must be formed nearly instantaneously after the collision
or must preexist in the incoming projectiles. They are therefore very
sensitive to the strong color fields present in the initial stage and
their correlations.

In a recent paper~\cite{DusliGLV1} (see also
~\cite{DumitJ3,GelisLV5}), it was shown in detail how these long range
correlations arise as a consequence of the saturation of gluons with
momenta $k_\perp \leq Q_\mathrm{s}$ in the nuclear wave functions 
\emph{before} the collision.  Here $Q_\mathrm{s} > \Lambda_{\rm QCD}$
is a semi-hard saturation scale~\cite{GriboLR1,MuellQ1}, which
provides a measure of the range of color correlations in a {\it
  nucleon or nucleus}. This scale is large for {\it either high
  energies or large nuclei or both}, suggesting the presence of strong
universal dynamical color correlations at distances smaller than the
scale of confinement dynamics $~1/\Lambda_{\rm QCD}$ in QCD. Because
$Q_\mathrm{s}$ is large, a quantitative understanding of the highly
non-perturbative dynamics of saturated gluons in feasible in weak
coupling, and is realized in the Color Glass Condensate (CGC)
effective theory~\cite{IancuLM3,IancuV1,GelisIJV1} of hadron and
nuclear wavefunctions at high energies. The reference~\cite{DusliGLV1}
provides a quantitative comparison with the PHOBOS data and makes
predictions for similar correlations in A+A collisions at the LHC.

When two hadrons or nuclei collide at high energies, the CGC framework
predicts that strong longitudinal chromo-electric and chromo-magnetic
fields are produced that are nearly
boost-invariant~\cite{KovneMW1,KharzKV1}; this form of matter has been
called the Glasma~\cite{LappiM1}. We should emphasize here at the
outset that the properties of the Glasma are not computed in an
\emph{ad hoc} model but follow from high energy factorization theorems
which relate multi-particle dynamics in the Glasma to multi-parton
correlations in saturated hadron/nuclear
wavefunctions~\cite{GelisLV3,GelisLV4,GelisLV5}. A key feature of the
Glasma is that the transverse correlation length of these strong
longitudinal chromo-electric and chromo-magnetic fields is
$1/Q_\mathrm{s}$.  Two-particle~\cite{DumitGMV1,GavinMM1,LappiSV1} and
three-particle~\cite{DusliFV1} correlations were computed in this
Glasma ``flux tube'' picture. We should note that the $n$-particle
correlations can be shown to satisfy a negative binomial
distribution~\cite{GelisLM1}--providing a microscopic derivation of
this widely used multiplicity distribution.  These correlations are
independent of $\Delta \eta$ up to quantum corrections that become
important for $\Delta\eta \gtrsim 1/\alpha_\mathrm{s}$. In heavy ion
collisions the collimation in $\Delta \phi$ comes primarily from the
subsequent radial flow of the correlated
particles~\cite{Volos1,Shury2,PruneGV1,Takahashi:2009na,Moschelli:2009tg,Werner:2010aa}.

However, as discussed in more detail below, an intrinsic $\Delta \phi$
collimation is also present independently of the effects from flow in
the later stages of the collision.  It is very weak for the small
$Q_\mathrm{s}$ values that are attained in p+p, d+A and peripheral A+A
collisions at RHIC. At LHC energies, for the large multiplicity cuts
performed by the CMS experiment (which select central impact
parameters), the signal is stronger because $Q_\mathrm{s}$ is larger
but not as strong as in central $A+A$ collisions, because the
effect of radial flow is smaller or absent.  Indeed, ``predictions''
previous to the CMS announcement existed~\cite{AD_rikenwkshp} but were
not submitted for publication because it appeared inconceivable that
such a small signal would be detected experimentally--this places the
{\it tour de force} measurement by CMS in perspective!

We shall now show some of our results that exhibit the same
qualitative features as the CMS results. More quantitative comparisons
would require a more detailed understanding of the impact parameter
dependence affecting the normalization, experimental efficiency and
acceptance effects and the contribution of radial flow.  We hope to be
able to make more detailed comparisons in the near future.

\begin{figure}[tbh]
\begin{center}
\includegraphics[width=0.4\textwidth]{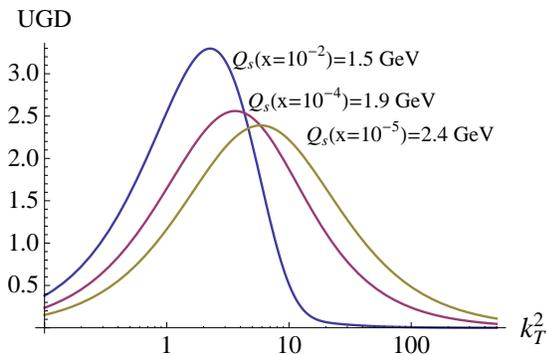}
\end{center}
\caption{\label{fig:ugd} The unintegrated gluon
  distribution as a function of transverse momentum squared for 
 running coupling for three values of the momentum fraction $x$.
}
\end{figure}

\section{Long range two-particle correlation}

The strength of two-particle correlations is conveniently represented
by the function
\begin{equation}
C_2(\pp, y_p, \qp, y_q) \equiv \frac{\ud N}{\ud y} 
\left[
 \frac{\frac{\ud N_2}{\ud^2\pp \ud y_p \ud^2\qp \ud y_q} }
{\frac{\ud N}{\ud^2 \pp \ud y_p }  \frac{\ud N}{\ud^2\qp \ud y_q} }
-1
\right],
\label{eq:defc2}
\end{equation}
The expression in eq.~(\ref{eq:defc2}) corresponds
(qualitatively) to the quantity $R(\Delta\phi,\Delta \eta)$ plotted in
the CMS paper.  Our result for the correlated two-particle
distribution in eq.~(\ref{eq:defc2}) can be expressed
as~\cite{DusliGLV1} \setlength\arraycolsep{.4pt}
\begin{eqnarray}
\label{eq:double-inclusive}
\frac{\ud N_2}{\ud^2\pp \ud y_p \ud^2\qp \ud y_q}
&=&
\frac{\alpha_s^{2}}{16 \pi^{10}}
\frac{N_c^2 S_\perp}{(N_c^2-1)^3\; \pp^2\qp^2} 
\\ & & \hspace{-2.5cm}
\times
\int \ud^2\kp 
 \Big\{ 
\Phi_{A}^2(y_p,\kp)\, \Phi_{B}(y_p,\pp-\kp)
\nonumber \\
& & \hspace{-1.5cm} \times \left[
\Phi_{B}(y_q,\qp+\kp)
+
\Phi_{B}(y_q,\qp-\kp)
\right]
\nonumber\\
&& \hspace{-2cm} + \Phi_{B}^2(y_q, \kp)\, \Phi_{A}(y_p,\pp-\kp)
\nonumber \\
& & \hspace{-1.5cm}\times \left[
\Phi_{A}(y_q,\qp+\kp)
+
\Phi_{A}(y_q,\qp-\kp)
\right]\!
\Big\}\, .
\nonumber
\end{eqnarray}
Here the $\Phi$'s are \emph{unintegrated gluon distributions} (UGD) per unit of
transverse area in projectiles $A$ and $B$ and $S_\perp$ is the transverse
overlap area of the two hadrons. 
Equation \ref{eq:double-inclusive} is
based on the formalism developed in~\cite{GelisLV5,DumitJ3}
and is derived in ref.~\cite{DusliGLV1}.
The single inclusive gluon spectrum can be expressed in the same
notation as~\cite{KovchM3,Braun1,FujiiGV2}
\begin{multline}
\frac{\ud N}{\ud^2\pp \ud y_p}
=\frac{\alpha_s N_c S_\perp}{\pi^4 (N_c^2-1)}\frac{1}{\pp^2}
\\ \times
\int \frac{\ud^2\kp}{(2\pi)^2} \Phi_{A}(y_p,\kp)\, \Phi_{B}(y_p,\pp-\kp)\; .
\label{eq:single-inclusive}
\end{multline}
The above expressions (eqs.~\ref{eq:double-inclusive} and
\ref{eq:single-inclusive}) are valid 
to leading logarithmic accuracy in $x$ and for momenta 
$p_\perp,q_\perp \gtrsim Q_\mathrm{s}$. 

The important ingredient in the above expressions is the 
unintegrated gluon density which is a universal quantity and can be
constrained from fits to DIS on hadron~\cite{AlbacAMS1} and nuclear
targets~\cite{DusliGLV1}--these unintegrated gluon densities have been
used to compute single inclusive and double inclusive distributions in
d+Au collisions at RHIC~\cite{AlbacM1}.  The only difference between
different targets are in the initial conditions determined at an
initial $x_0$--chosen to be $x_0=0.01$. The evolution of the UGD with
rapidity is controlled by the Balitsky--Kovchegov (BK)
equation~\cite{Balit2,Kovch2}, which is a non--linear evolution
equation describing both gluon emission and multiple scattering
effects. We note that our unintegrated gluon distributions have
recently been shown to be equivalent to those derived in the framework
of transverse momentum dependent (TMD) parton
distributions~\cite{DominXY1}. In fig.~\ref{fig:ugd} we show the
structure of this unintegrated gluon distribution and its evolution
with rapidity for a running coupling. The running of the coupling
slows the evolution significantly which is essential to describe DIS
and hadron scattering data.

\section{The ridge in proton proton collisions}

In proton--proton collisions, unlike nucleus-nucleus collisions, it is
not clear that there can be a sufficient amount of transverse flow
to provide the collimation around $\Delta\phi\approx 0$
necessary to explain the ridge. Blast wave fits to proton
single particle spectra at low $p_\perp$ are suggestive of
non--vanishing radial flow, and this in principle could provide an
additional collimation of the signal. We also note that in the
highest multiplicity proton--proton events the charged particle
multiplicity per unit rapidity is comparable to that found in
semi--central Cu--Cu collisions.  The possibility that additional
collimation of the signal due to flow might be needed to reproduce the
full strength of the measured correlation cannot be completely ruled
out. However, the absence of a correlation signal for small transverse
momenta in proton-proton collisions (contrary to the nucleus-nucleus
case) suggests that hydrodynamical flow would not be the dominant
contribution here.

\begin{figure}[htb]
\begin{center}
\includegraphics[width=0.25\textwidth]{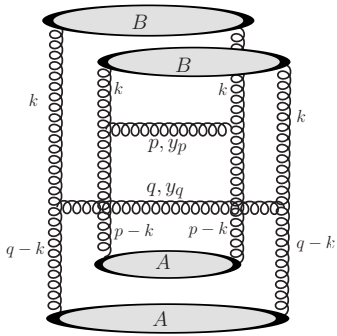}
\end{center}
\caption{\label{fig:diag}
 A typical diagram which gives an angular collimation.}
\end{figure}
The main point of this paper is that there is an intrinsic angular
correlation (in addition to the long range rapidity correlation)
coming from the particle production process on transverse distance
scales $1/Q_\mathrm{s}$ much smaller than the proton size. One of the
diagrams that gives a collimation for $\Delta \phi\approx 0$ is shown
in fig.~\ref{fig:diag}. There is only a single loop momentum $\kp$ in
these two-particle production diagrams, as opposed to uncorrelated
production. The transverse momenta flowing into the ``blobs'' for
hadron~A in this diagram are $|\pp-\kp|$ and $|\qp-\kp|$,
respectively. If we combine this with the fact that the unintegrated
distributions peak about $Q_\mathrm{s}$ (fig.~\ref{fig:ugd}) it
follows that the \emph{largest} contribution is obtained when both
$|\pp-\kp|\sim Q_\mathrm{s}$ and $|\qp-\kp|\sim Q_\mathrm{s}$. These
conditions are satisfied simultaneously, for one and the same $\kp$,
if $\pp$ and $\qp$ are parallel, leading to angular collimation. The
scale of the angular dependence is ${\cal O}(1)$.  Also, if
parametrically $p_\perp$ and $q_\perp$ are much smaller or much
greater than $Q_\mathrm{s}$, the collimation disappears.

In contrast, in collinear factorization, at leading order both gluons
are produced from the same ladder leading to collimation about
$\Delta\phi\sim\pi$ (``dijet''). One may therefore expect that
diagrams such as fig.~\ref{fig:diag} give a large contribution to
$C(\pp,y_p,\qp,y_q)$ when $\Delta\phi\ll\pi$ and $|y_p-y_q|\gsim 1$
(see, also, discussion in ref.~\cite{DumitJ3}).

\begin{figure*}
\begin{center}
\includegraphics[width=0.8\textwidth]{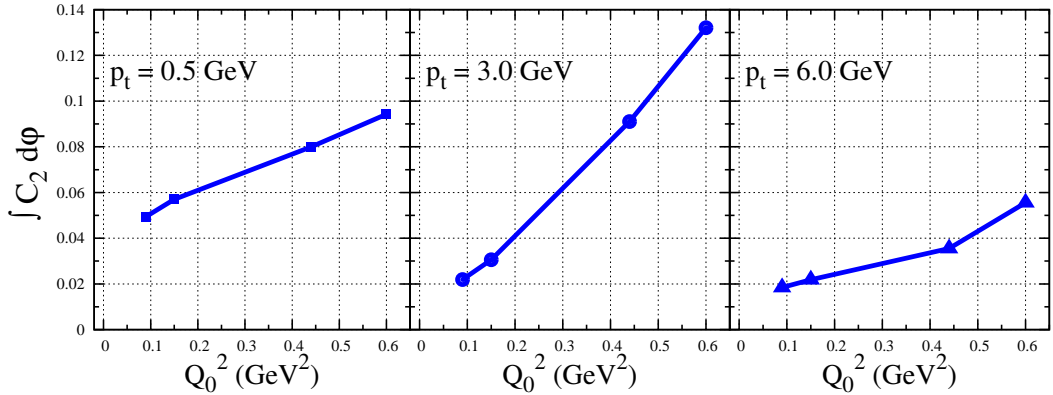}
\end{center}
\caption{The correlation function $C_2$  for different $p_\perp$
as a function of  $Q_0^2$, the saturation scale at the initial 
condition of the evolution $x=x_0$. Increasing values of
$Q_0$ correspond to smaller impact parameters and larger
total charged multiplicities.}
\label{fig:five}
\end{figure*}

\begin{figure}[tbh]
\begin{center}
\includegraphics[width=0.48\textwidth]{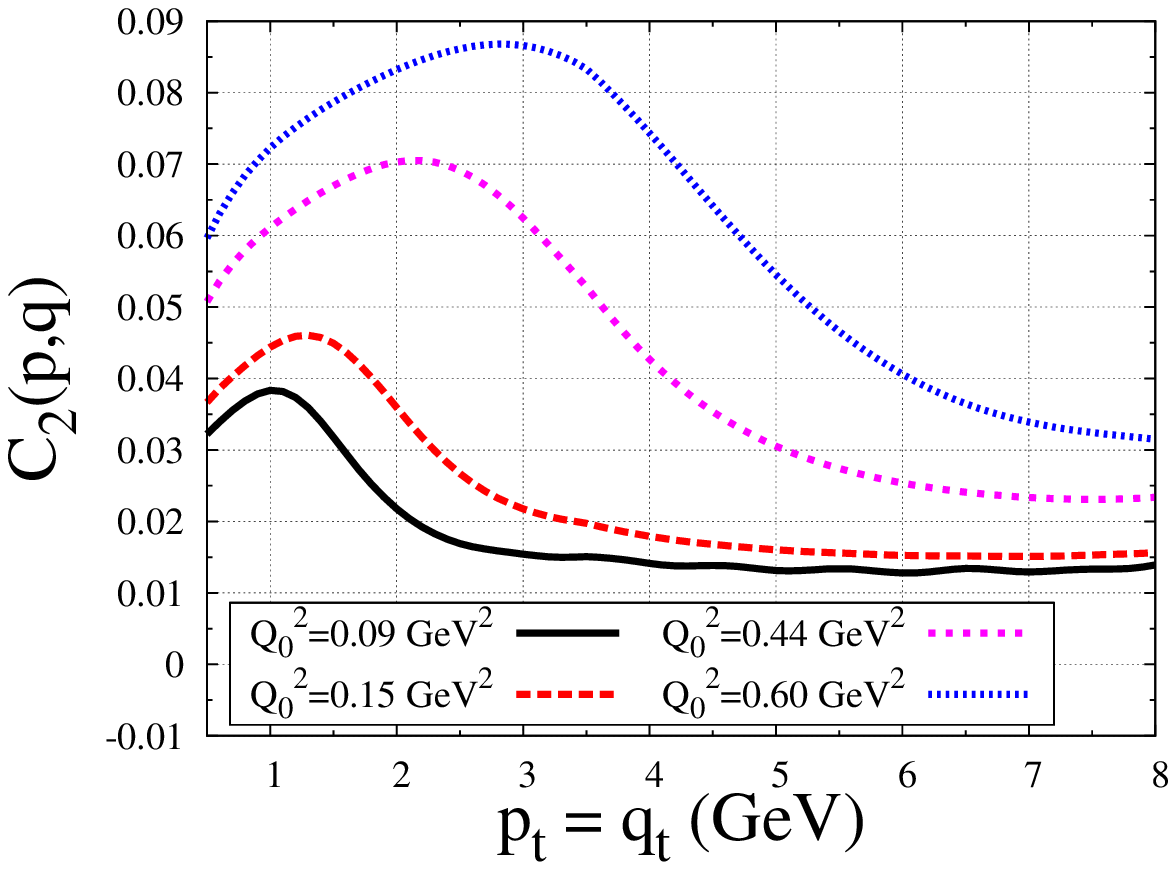}
\includegraphics[width=0.48\textwidth]{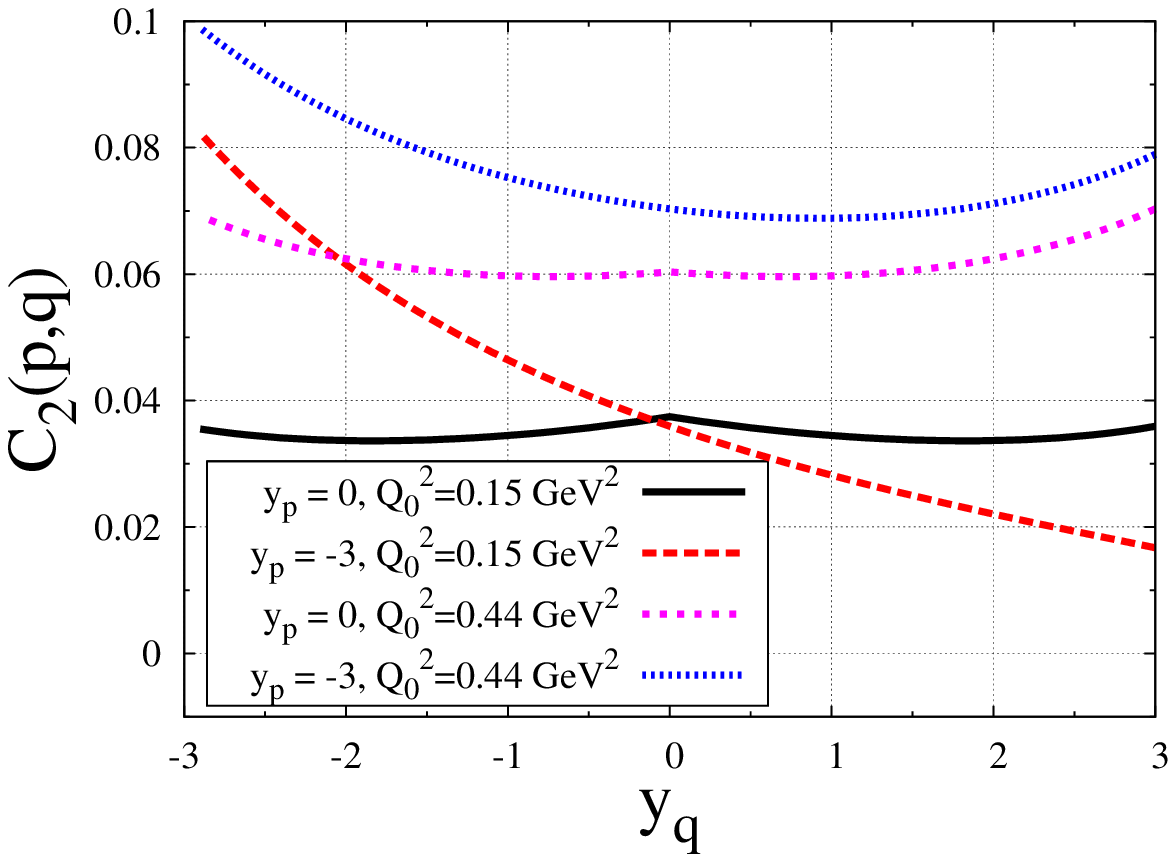}
\end{center}
\caption{\label{fig:six} Top: $p_\perp$ dependence of $C_2$ at $\Delta
  \phi\approx 0$ for four different initial scales ranging from
  minimum bias events to high multiplicity events.  Bottom: Rapidity
  dependence of $C_2$ at $\Delta\phi= 0$ for $p_\perp=q_\perp = 2$~GeV
  as a function of $y_q$, for two different initial scales and $y_p$.
  See text for discussion.}
\end{figure}
Our results are summarized in figs.~\ref{fig:five} and~\ref{fig:six}.
In fig.~\ref{fig:five} we show $C_2(\pp, y_p, \qp, y_q)$ for $p_\perp
\approx q_\perp$ as a function of $Q_0^2$, the saturation scale at the
initial condition of the evolution $x=x_0$.  A larger saturation scale
corresponds to a more central collision and thus to a higher
multiplicity. Thus our result shown in fig.~\ref{fig:five} agrees with
the CMS observation of the ridge correlation becoming visible for
higher multiplicity events. The exact relation between the value of
$Q_\mathrm{s}$ and the observed multiplicity depends on the overlap
area of the two protons and on the impact parameter dependence of the
saturation scale (see e.g.~\cite{KowalMW1,Levin:2010zy}).
Establishing the normalization without additional free parameters
would require going beyond the $k_\perp$-factorized expression
(\ref{eq:single-inclusive}) for the single inclusive
spectrum~\cite{Blaizot:2010kh}.  However, in the appropriately
normalized correlation (\ref{eq:defc2}) these factors cancel and our
result does not depend strongly on the size of the overlap area.  The
result shown in fig.~\ref{fig:five} is obtained by integrating over
$\Delta \phi$ for $0\leq \Delta \phi \leq \pi/2$.  For a fixed
$p_\perp$, we see that the correlation increases with centrality. The
systematics is striking. For all centralities, the collimation is weak
for low $p_\perp$. For $p_\perp > 1$~GeV, the correlation grows but at
$p_\perp > 6$~GeV, the correlation decreases below the value at
$p_\perp=0.5$~GeV and is smaller for all higher $p_\perp$ values. The
reader should keep in mind that this result is for gluons.
Fragmentation effects may approximately cancel in the $C_2$ ratio but
the $p_\perp$ of the hadrons is of course lower than that of the
gluons.  While this plot cannot be compared directly to
fig.~\ref{fig:one} (bottom) it shows a non trivial systematics very
similar to it.

In fig.~\ref{fig:six} (top) we plot the $p_\perp$ dependence of
$C_2$. We see that the distribution is peaked approximately at the
value of $Q_\mathrm{s}$ evolved to the rapidities of interest and that
the correlations drop off sharply at large $p_\perp$. For low
$p_\perp$, one has less theoretical control on how the distributions
fall off but the trend is unmistakable. In fig.~\ref{fig:six}
(bottom), we show the rapidity dependence for fixed $p_\perp= q_\perp=2$~GeV
as a function of $y_q$, given two different $y_p$ and two different
values of $Q_0$.  For $y_p=0$, we see that the distributions are
symmetric around $y_q=0$. This is not the case for $y_p=-3$, because
QCD evolution (gluon radiation) between the projectiles to the
produced gluons is asymmetric. The systematics could in principle be
investigated to test our predictions in detail.

\section{Summary}

The physics of gluon saturation opens up a novel domain of many body
QCD at high energies and may contain many surprises which provide
deeper insight into the fundamental properties of strong interactions.
One of these is the ridge in proton-proton collisions and it is likely
that many more will be revealed at the LHC.  The properties of
saturated gluons, in particular the evolution of multi-parton
correlations with energy, can be computed in the Color Glass
Condensate effective field theory. High energy factorization theorems
allow us to relate these multi-parton correlations in the hadron and
nuclear wavefunctions to inclusive multi-parton final states in
collisions. The key expression eq.~(\ref{eq:double-inclusive}) is one
of the results of this approach. It provided a detailed understanding
of the ridge in A+A collisions at RHIC. It predicts a ridge in p+p
collisions.

We computed in this paper the two-particle correlation $C_2$
(eq.~(\ref{eq:defc2})) and showed that it shares the same key features
as the quantity $R$ measured by the CMS experiment:
\begin{itemize}
\item It is long range in rapidity, and exhibits a collimation for
  high multiplicity events for a narrow window of a few GeV in the
  transverse momenta of the pair particles. The effect becomes
  systematically weaker both below and above the kinematic window as
  seen in the CMS result.
\item The two-particle correlation has the same strength for both like
  and unlike sign pairs. This feature was seen already at RHIC and is
  consistent with gluon emission from a source that is (nearly)
  uniform in rapidity, ``the glasma flux tube''. It is not consistent
  with emission from jets.
\item The two-particle correlation is relatively flat in the $\eta_1$
  versus $\eta_2$ plane. Again, this feature is natural for gluon
  emission from nearly boost invariant classical sources. We predict a
  similarly flat signal strength in three-particle
  correlations~\cite{DusliFV1}.
\end{itemize}
Correlations of trigger particles at non central rapidities with
associated particles at other rapidities can provide sensitive tests
of this picture.  In order to make a quantitive comparison one
needs to add the short range jet component (which can be generated with
one of the available event--generators) to our long range correlation.
After including fragmentation one could then form the same quantities
as measured in experiment. This is left for further study.

\section*{Acknowledgements}
We would like to acknowledge James (Bj) Bjorken who in 
conversations with us at BNL and at the INT
in May 2010 was enthusiastic about the prospects of  
observing the ridge in pp collisions.
We gratefully acknowledge useful
conversations with Larry McLerran, Robert Pisarski and Nick Samios. We
thank Gunther Roland and David Wei Li of the CMS collaboration for
communications regarding the experimental findings. A.D. and J.J-M.
gratefully acknowledge support by the DOE Office of Nuclear Physics
through Grant No.~DE-FG02-09ER41620 and from The City University of
New York through the PSC-CUNY Research Award Program, grant
63382-00~41.  K.D. and R.V. were supported by by the US Department of
Energy under DOE Contract No.~DE-AC02-98CH10886. T.L. is supported by
the Academy of Finland, project 126604. F.G. is supported in part by
Agence Nationale de la Recherche via the program
ANR-06-BLAN-0285-01. T.L. and F.G. thank the Institute for Nuclear
Theory at the University of Washington for partial support during the
completion of this work.

\end{document}